# Bound-states for generalized trigonometric and hyperbolic Pöschl-Teller potentials

A. D. Alhaidari[a], I. A. Assi[b], and A. Mebirouk[c]

[a] *Saudi Center for Theoretical Physics, P.O. Box 32741, Jeddah 21438, Saudi Arabia*

[b] *Department of Physics and Physical Oceanography, Memorial University of Newfoundland, St. John's, Newfoundland & Labrador, A1B 3X7, Canada*

[c] *Mathematical Modeling and Numerical Simulation Laboratory, Badji Mokhtar University, BP 12, Annaba, Algeria*

**Abstract:** We use the "tridiagonal representation approach" to solve the time-independent Schrödinger equation for the bound states of generalized versions of the trigonometric and hyperbolic Pöschl-Teller potentials. These new solvable potentials do not belong to the conventional class of exactly solvable problems. The solutions are finite series of square integrable functions written in terms of the Jacobi polynomial.

## 1. Introduction

In quantum mechanics, potential models are used to describe the behavior of physical systems as they interact with their environment. The success of a model depends on how well it represents the system and predicts its behavior in a given scenario. In principle, there is an infinite number of such potential models. However, the most relevant ones are those that could be solved exactly since then one can explore all physical scenarios that may arise, especially at critical couplings to the environment where the system makes a dramatic transition from one phase to another. Moreover, exact solutions can easily give access to the most significant properties of the system such as the energy spectrum of bound states, phase shift of the scattering states, thermal properties, density of states, etc. On the other hand, it is worth mentioning that the class of exactly solvable potentials is very small. The most well-known members of this class include the Coulomb, harmonic oscillator, Morse, Pöschl-Teller, Eckart, and Scarf potentials [1,2]. Many researchers over the years (present company included) devised different methods in an attempt to find exact solutions of the wave equation (e.g., Schrödinger and Dirac) for new potential models. If and when successful, only very few discovered potentials join the class of exactly solvable potentials. The most popular methods of solution of the wave equation include factorization, supersymmetry, path integral, point canonical transformation, shape invariance, Darboux transformation, asymptotic iteration, and the Nikiforov-Uvarov method. Our choice is the Tridiagonal Representation Approach (TRA). It is an algebraic method for solving linear ordinary differential equations of the second order [3-5].

For a detailed description of the method and how it is used to solve quantum mechanical problems, one may consult [6,7] and references therein.

In this work, we employ the TRA to obtain exact bound-state solution of the time-independent one-dimensional Schrödinger equation for the following four-parameter potential models

$$V_I(x) = \frac{V_0}{\sinh^4(\kappa x)} + \frac{A}{\sinh^2(\kappa x)} + \frac{B}{\cosh^2(\kappa x)}, \qquad x \geq 0. \tag{1a}$$

$$V_{II}(x) = \frac{V_0}{\cos^4(\pi x/2a)} + \frac{C}{\cos^2(\pi x/2a)} + \frac{D}{\sin^2(\pi x/2a)}, \qquad 0 \leq x \leq a. \tag{1b}$$

Without the most singular $V_0$-term, they become the hyperbolic and trigonometric Pöschl-Teller potentials, respectively [2,8]. All potential parameters are real with $V_0 > 0$ and $D > 0$. Under the transformation $V_I(x) \mapsto V_I(-x)$ and $V_{II}(x) \mapsto V_{II}(-x)$, these potentials get reflected on the $x = 0$ axis but remain physically the same (isospectral). Moreover, the transformation $V_{II}(x) \mapsto V_{II}(x \pm a)$ interchanges the trigonometric functions as $\cos \leftrightarrow \sin$ and results in a horizontal displacement of the potential along the $x$-axis by $\mp a$ but also remains isospectral. Consequently, the potential $\tilde{V}_{II}(x) := V_{II}(-x + a)$,

$$\tilde{V}_{II}(x) = \frac{V_0}{\sin^4(\pi x/2a)} + \frac{C}{\sin^2(\pi x/2a)} + \frac{D}{\cos^2(\pi x/2a)}, \qquad 0 \leq x \leq a. \tag{1c}$$

is isospectral to $V_{II}(x)$. Moreover, it is easy to see that the complex transformation $x \mapsto ix$ in the Schrödinger equation with this potential changes it to that for $-V_I(x)$ with $C \mapsto -A$, $D \mapsto B$ and the energy $E \mapsto -E$. This stands to show that complex extension of quantum mechanics [9] and doing quantum mechanics on complex coordinates provide a unified treatment for potentials that seemingly have different physical characteristics.

Now, both potentials (1a) and (1b) do not belong to the conventional class of exactly solvable quantum mechanical problems. Potential $V_I(x)$ is inverse-quartic and inverse-square singular at the origin with singularity strength of $V_0$ and $A$, respectively. Moreover, it is short-range with $\kappa$ being the range measure, which is a positive scale parameter of inverse length dimension. On the other hand, $V_{II}(x)$ is a potential well of width $a$ with an inverse-square singularity at the left wall of strength $D$. However, on the right wall, it is inverse-quartic and inverse-square singular with singularity strength of $V_0$ and $C$, respectively. It is obvious that the energy spectrum for $V_{II}(x)$ is discrete, infinite, and bounded from below. Whereas, with a proper choice of values of the parameters, $V_I(x)$ could support a finite number of bound states in addition to a resonance structure. These proper parameter values are required to produce one or two positive real roots for the cubic equation that results from the condition $(dV_I/dx)_{x_0} = 0$. A necessary, but not sufficient, condition for the existence of bound states is that $V_I(x_0) < 0$. The existence of two different real positive roots of the cubic equation implies the possibility of resonance. This cubic equation, which is associated with $V_I(x)$, reads as follows

$$(A+B)s^3 + 2(V_0 + A)s^2 + (4V_0 + A)s + 2V_0 = 0, \tag{2}$$

where $s = \sinh^2(\kappa x)$. Since the constant term in this equation is positive, then Descartes' rule of signs [10] dictates that the existence of one positive real root requires that the coefficient of the $s^3$ term be negative. That is, a necessary (may not be sufficient) condition for bound states without resonance is $(A+B) < 0$. Figure 1 shows several plots of the two potential functions (in units of $V_0$) for several values of the associated parameters. Figure 2 is the spectral phase diagram (SPD) for $V_I(x)$ showing the distribution of the energy spectrum (scattering states, bound states, and resonances) as a function of the potential parameters. The SPD for potential systems was introduced in Ref. [11] and is explained here briefly in section B.3 of Appendix B.

In sections 2 and 3, we formulate the problem within the TRA and solve it for the two potential models, respectively. We conclude in section 4 with some remarks and a discussion of our main findings.

## 2. TRA solution of the hyperbolic potential

We start by writing the solution of the Schrödinger equation, $\mathcal{D}\psi(x) = 0$, as a bounded convergent series of discrete square-integrable functions $\{\phi_n\}$. That is, $\psi(x) = \sum_n f_n \phi_n(y)$, where $y = y(x)$ is a coordinate transformation and $\{f_n\}$ are expansion coefficients. The basis set $\{\phi_n\}$ must be complete and should result in a tridiagonal matrix representation for the wave operator, $\langle \phi_n | \mathcal{D} | \phi_m \rangle$. That is, the action of the wave operator on the basis element should read

$$\mathcal{D}\phi_n(y) = W(y)\left[g_n \phi_n(y) + c_{n-1}\phi_{n-1}(y) + d_n \phi_{n+1}(y)\right], \tag{3}$$

where $W(y)$ is a node-less entire function and $\{g_n, c_n, d_n\}$ are constant coefficients. Hence, the wave equation, $\mathcal{D}\psi(x) = 0$, becomes a three-term recursion relation for the expansion coefficients $\{f_n\}$ that reads

$$g_n F_n + d_{n-1} F_{n-1} + c_n F_{n+1} = 0, \tag{4}$$

where we wrote $f_n = f_0 F_n$ making $F_0 = 1$. Consequently, the solution of the wave equation reduces to an algebraic solution of the discrete relation (4) and the set $\{f_n\}$ contains all physical information about the system modelled by the potential.

Now, to solve for the continuous spectrum or for an infinite discrete spectrum of a given system, completeness of the set $\{\phi_n\}$ implies that it is an infinite set. However, for systems with a finite number of bound states (e.g., the system associated with the potential $V_I$), a finite basis set $\{\phi_n\}_{n=0}^N$ could produce a faithful representation of the physical system provided that the number of bound states is less than or equal to the size of the basis $N+1$. Additionally, for quasi-exact solution where one looks for a finite portion of the infinite discrete spectrum (e.g., the system associated with the potential $V_{II}$), such finite basis set could lead to a very good approximation for that portion of the spectrum with an accuracy that increases with $N$.

For the system modelled by either $V_I(x)$ or $V_{II}(x)$, we choose a finite basis set with the following elements

$$\phi_n(y) = (y-1)^\alpha (y+1)^{-\beta} Q_n^{(\mu,\nu)}(y), \tag{5}$$

where $Q_n^{(\mu,\nu)}(y)$ is the finite Jacobi polynomial defined on the semi-infinite real line $y(x) \geq 1$ as shown in Appendix A. The real basis parameters $\{\alpha, \beta, \mu, \nu\}$ are to be determined below in terms of the physical parameters by the TRA constraints. Moreover, $n = 0, 1, .., N$ with $N = \lfloor -\frac{\mu+\nu+1}{2} \rfloor$ where $\lfloor z \rfloor$ stands for the largest integer less than $z$.

In the atomic units $\hbar = m = 1$, the time-independent Schrödinger equation, $\mathcal{D}\psi(x) = 0$, in the configuration space $x$ for the potential $V_I(x)$ and energy $E$ reads as follows:

$$\left[ -\frac{1}{2}\frac{d^2}{dx^2} + V_I(x) - E \right]\psi(x) = 0. \tag{6}$$

We choose the coordinate transformation $y(x) = \frac{2}{\tanh^2(\kappa x)} - 1$. Writing the differential operator $\frac{d^2}{dx^2}$ and the potential $V_I(x)$ in terms of the dimensionless variable $y$, Eq. (6) becomes

$$\mathcal{D}\psi(x) = -\kappa^2 (y-1) \left[ (y^2-1)\frac{d^2}{dy^2} + \frac{1}{2}(3y+1)\frac{d}{dy} \right.$$
$$\left. -\frac{B/\kappa^2}{y+1} - \frac{V_0 - A}{4\kappa^2}(y-1) - \frac{A}{4\kappa^2}(y+1) + \frac{E/\kappa^2}{y-1} \right]\psi(x) = 0 \tag{7}$$

To solve this equation, we substitute the series $\psi(x) = \sum_n f_n \phi_n(y)$. Consequently, we need to evaluate the action of the wave operator on the basis elements, $\mathcal{D}|\phi_n\rangle$, and then impose the TRA constraint (3). To achieve that we choose the basis parameters $2\alpha = \mu$ and $2\beta = -\nu - \frac{1}{2}$ and use the differential equation of the Jacobi polynomial $Q_n^{(\mu,\nu)}(y)$ given in Appendix A by (A2) to obtain

$$\mathcal{D}|\phi_n(y)\rangle = -\frac{\kappa^2}{2}\frac{(y-1)^{\alpha+1}}{(y+1)^\beta}\left\{ \frac{\mu^2}{y-1} - \frac{\nu^2 - \frac{1}{4}}{y+1} + \frac{1}{2}\left[(2n+\mu+\nu+1)^2 - \frac{1}{4}\right] \right.$$
$$\left. + \frac{\varepsilon}{y-1} - \frac{2B/\kappa^2}{y+1} - \frac{V_0}{2\kappa^2}y + \frac{V_0 - 2A}{2\kappa^2} \right\} Q_n^{(\mu,\nu)}(y) \tag{8}$$

where $\varepsilon = 2E/\kappa^2$. Now, the TRA constraint (3) and the recursion relation of the Jacobi polynomials (A3) dictate that terms inside the curly brackets in (8) must be linear in $y$. Thus, we should choose the Jacobi polynomial parameters as

$$\mu^2 = -\varepsilon, \qquad \nu^2 = \frac{1}{4} - (2B/\kappa^2). \tag{9}$$

Reality dictates that $E < 0$ and $B \leq \kappa^2/8$, which means that the TRA solution obtained here is valid only for the negative energy bound states. Moreover, the basis (5) becomes energy dependent via the parameter $\mu$. The polynomial parameters inequalities, $\mu > -1$ and $\mu + \nu < -2N - 1$, dictate that $\nu = -\sqrt{\frac{1}{4} - (2B/\kappa^2)}$ and $\mu = \sqrt{-\varepsilon}$. With this choice of basis parameters, Eq. (8) becomes

$$\mathcal{D}|\phi_n\rangle = -\frac{\kappa^2}{4} \frac{(y-1)^{\alpha+1}}{(y+1)^\beta} \left\{ \left[ (2n+\mu+\nu+1)^2 - \frac{1}{4} \right] + \frac{V_0 - 2A}{\kappa^2} - \frac{V_0}{\kappa^2} y \right\} Q_n^{(\mu,\nu)}(y). \tag{10}$$

Using the three-term recursion relation for the Jacobi polynomials (A3) in this equation and comparing the result to the TRA constraint (3), we obtain

$$W(y) = \frac{V_0}{4}(y-1), \tag{11a}$$

$$g_n = \left\{ \left[ \left(n + \frac{\mu+\nu+1}{2}\right)^2 - \frac{1}{16} \right] + \frac{V_0 - 2A}{4\kappa^2} \right\} \frac{4\kappa^2}{-V_0} + \frac{\nu^2 - \mu^2}{(2n+\mu+\nu)(2n+\mu+\nu+2)}, \tag{11b}$$

$$c_n = \frac{2(n+\mu+1)(n+\nu+1)}{(2n+\mu+\nu+2)(2n+\mu+\nu+3)}, \quad d_n = \frac{2(n+1)(n+\mu+\nu+1)}{(2n+\mu+\nu+1)(2n+\mu+\nu+2)}. \tag{11c}$$

For $n = 0, 1, .., N$ and $N = \left\lfloor -\frac{\mu+\nu+1}{2} \right\rfloor$, we can show that $c_n d_n > 0$ for all $0 \leq n \leq N - 1$. Then according to Favard theorem [12] (a.k.a. the spectral theorem; see section 2.5 in [13]) the sequence $\{F_n(z)\}$ satisfying the three-term recursion relation (4) will form a set of orthogonal polynomials with $f_0^2(z)$ as their positive definite weight function [7,14,15] and where $z$ is some proper parameter that depends on the potential parameters $V_0$ and $A$. If we define the polynomial

$$P_n = \frac{(\mu+1)_n (\nu+1)_n}{n!(\mu+\nu+1)_n} \frac{\mu+\nu+1}{2n+\mu+\nu+1} F_n := G_n F_n, \tag{12}$$

Then the recursion relation (4) written for $\{P_n\}$ becomes identical to that of the polynomial $H_n^{(\mu,\nu)}(z;\gamma,\theta)$ shown in Appendix A as Eq. (A7) with the following parameters and argument

$$\gamma^2 = 1/16, \quad \cos\theta = 1 - \frac{2A}{V_0}, \quad z^2 = \frac{4\kappa^4}{A(V_0 - A)}. \tag{13a}$$

The condition $|\cos\theta| \leq 1$ dictates that $0 < A < V_0$, which is the same condition that guarantees reality of $z$. On the other hand, comparing the recursion relation (4) written for $\{P_n\}$ to Eq. (A8), we conclude that $P_n = \tilde{H}_n^{(\mu,\nu)}(z;\gamma,\theta)$ with

$$\gamma^2 = 1/16, \quad \cosh\theta = 1 - \frac{2A}{V_0}, \quad z^2 = \frac{4\kappa^4}{A(A - V_0)}. \tag{13b}$$

The condition $\cosh\theta \geq 1$ dictates that $A < 0$, which is the same condition that guarantees reality of $z$. Finally, the $m^{\text{th}}$ bound state wavefunction with energy $E_m = \frac{1}{2}\kappa^2 \varepsilon_m$ is written as one of the following two finite series depending on the range of values of the potential parameter $A$

$$\psi_m(x) = \left(\sqrt{2}\right)^{\mu_m+\nu+\frac{1}{2}} f_0(z)[\cosh(\kappa x)]^{\nu+\frac{1}{2}}[\sinh(\kappa x)]^{-\mu_m-\nu-\frac{1}{2}}$$
$$\times \sum_{n=0}^{N_m} G_{n,m}^{-1} H_n^{(\mu_m,\nu)}\left(z;\tfrac{1}{4},\theta\right) Q_n^{(\mu_m,\nu)}(y), \qquad 0 < A < V_0 \tag{14a}$$

$$\psi_m(x) = \left(\sqrt{2}\right)^{\mu_m+\nu+\frac{1}{2}} f_0(z)[\cosh(\kappa x)]^{\nu+\frac{1}{2}}[\sinh(\kappa x)]^{-\mu_m-\nu-\frac{1}{2}}$$
$$\times \sum_{n=0}^{N_m} G_{n,m}^{-1} \tilde{H}_n^{(\mu_m,\nu)}\left(z;\tfrac{1}{4},\theta\right) Q_n^{(\mu_m,\nu)}(y), \qquad A < 0 \tag{14b}$$

where $\mu_m = \sqrt{-\varepsilon_m}$, $\nu = -\sqrt{\tfrac{1}{4}-(2B/\kappa^2)}$, $G_{n,m}$ is defined in (12) as $G_n(\mu_m)$, $z$ and $\theta$ are defined in (13a) or (13b), respectively. The integer $N_m$ is equal to $\left\lfloor \tfrac{1}{2}\left\{\sqrt{\tfrac{1}{4}-(2B/\kappa^2)} - \sqrt{-\varepsilon_m} - 1\right\} \right\rfloor$.

Now, the physical properties of the system are obtained from those of the associated TRA polynomials $H_n^{(\mu,\nu)}(z;\gamma,\theta)$ and $\tilde{H}_n^{(\mu,\nu)}(z;\gamma,\theta)$. Unfortunately, the complete analytic properties of these TRA polynomials are not yet known. It remains an open problem in orthogonal polynomials [16-18]. On the other hand, for a small number of terms in the wavefunction series (14), we can easily compute this polynomial explicitly for all these low degrees using its recursion relation and initial values. First, however, we need to obtain the energy spectrum. To do that, we use two independent numerical techniques as explained in Appendix B with results shown in Table 1 for two different sets of physical parameters. Once the bound state energies $\{E_m\}$ are known, the wavefunction is now fully determined by (14). Figure 3 shows the un-normalized wavefunctions corresponding to the energy spectrum given by Table 1.

## 3. TRA solution of the trigonometric potential

We repeat the same treatment in section 2 above for the potential model $V_{\text{II}}(x)$. However, we choose the following coordinate transformation and basis parameters

$$y(x) = 2\tan^2(\rho x)+1, \qquad 2\alpha = \mu + \tfrac{1}{2}, \qquad 2\beta = -\nu, \tag{15}$$

where $\rho = \pi/2a$. Consequently, the action of the wave operator on the basis elements becomes

$$\mathcal{D}\phi_n(y) = -\frac{\rho^2}{2}\frac{(y-1)^\alpha}{(y+1)^{\beta-1}}\left\{\frac{\mu^2-\tfrac{1}{4}}{y-1} - \frac{\nu^2}{y+1} + \frac{1}{2}\left[(2n+\mu+\nu+1)^2 - \frac{1}{4}\right]\right.$$
$$\left. + \frac{2E/\rho^2}{y+1} - \frac{2D/\rho^2}{y-1} - \frac{V_0+2C}{2\rho^2} - \frac{V_0}{2\rho^2}y\right\}Q_n^{(\mu,\nu)}(y) \tag{16}$$

where again we have used the differential equation of the Jacobi polynomial (A2). The TRA constraint (3) and the recursion relation of the Jacobi polynomials (A3) dictate that we assign the following values to the Jacobi polynomial parameters

$$\mu^2 = \tfrac{1}{4} + \left(2D/\rho^2\right), \qquad \nu^2 = \varepsilon, \tag{17}$$

where $\varepsilon = 2E/\rho^2$. Reality dictates that the energy spectrum is positive and $D \geq -\rho^2/8$. The latter condition is automatically satisfied since $D > 0$. The basis (5) becomes energy dependent via the parameter $v$. The polynomial parameters inequalities, $\mu > -1$ and $\mu + v < -2N - 1$, dictate that $\mu = \sqrt{\frac{1}{4} + (2D/\rho^2)}$ and $v = -\sqrt{\varepsilon}$. With these basis parameters, Eq. (16) becomes

$$\mathcal{D}|\phi_n\rangle = -\frac{\rho^2}{4}\frac{(y-1)^\alpha}{(y+1)^{\beta-1}}\left[(2n+\mu+v+1)^2 - \frac{1}{4} - \frac{V_0 + 2C}{\rho^2} - \frac{V_0}{\rho^2}y\right]Q_n^{(\mu,v)}(y). \tag{18}$$

Using the three-term recursion relation for the Jacobi polynomials (A3) in this equation and comparing the result to the TRA constraint (3), we obtain

$$W(y) = \frac{V_0}{4}(y+1), \tag{19a}$$

$$g_n = \left[\left(n + \frac{\mu+v+1}{2}\right)^2 - \frac{1}{16} - \frac{V_0 + 2C}{4\rho^2}\right]\frac{4\rho^2}{-V_0} + \frac{v^2 - \mu^2}{(2n+\mu+v)(2n+\mu+v+2)}, \tag{19b}$$

$$c_n = \frac{2(n+\mu+1)(n+v+1)}{(2n+\mu+v+2)(2n+\mu+v+3)}, \quad d_n = \frac{2(n+1)(n+\mu+v+1)}{(2n+\mu+v+1)(2n+\mu+v+2)}. \tag{19c}$$

Consequently, the recursion relation (4) written in terms of $\{P_n\}$ defined in (12) becomes identical to (A7) of the TRA polynomial $H_n^{(\mu,v)}(z;\gamma,\theta)$ with the following parameter and argument relations

$$\gamma^2 = 1/16, \quad \cos\theta = -[(2C/V_0) + 1], \quad z^2 = \frac{-4\rho^4}{C(C+V_0)}. \tag{20a}$$

The condition $|\cos\theta| \leq 1$ dictates that $-V_0 < C < 0$, which is the same condition that guarantees reality of $z$. On the other hand, comparing the recursion relation (4) written for $\{P_n\}$ to Eq. (A8), we conclude that $P_n = \tilde{H}_n^{(\mu,v)}(z;\gamma,\theta)$ with

$$\gamma^2 = 1/16, \quad \cosh\theta = -[(2C/V_0) + 1], \quad z^2 = \frac{4\rho^4}{C(C+V_0)}. \tag{20b}$$

The condition $\cosh\theta \geq 1$ dictates that $C < -V_0$, which is the same condition that guarantees reality of $z$. Finally, the $m^{\text{th}}$ bound state wavefunction with energy $E_m = \frac{1}{2}\rho^2\varepsilon_m$ is written as one of the following two finite series depending on the range of values of the potential parameter $C$

$$\psi_m(x) = \left(\sqrt{2}\right)^{\mu+v_m+\frac{1}{2}} f_0(z)[\sin(\rho x)]^{\mu+\frac{1}{2}}[\cos(\rho x)]^{-\mu-v_m-\frac{1}{2}}$$
$$\times \sum_{n=0}^{N_m} G_{n,m}^{-1} H_n^{(\mu,v_m)}\left(z;\tfrac{1}{4},\theta\right) Q_n^{(\mu,v_m)}(y), \quad 0 > C > -V_0 \tag{21a}$$

$$\psi_m(x) = \left(\sqrt{2}\right)^{\mu+v_m+\frac{1}{2}} f_0(z)[\sin(\rho x)]^{\mu+\frac{1}{2}}[\cos(\rho x)]^{-\mu-v_m-\frac{1}{2}}$$
$$\times \sum_{n=0}^{N_m} G_{n,m}^{-1} \tilde{H}_n^{(\mu,v_m)}\left(z;\tfrac{1}{4},\theta\right) Q_n^{(\mu,v_m)}(y), \quad C < -V_0 \tag{21b}$$

where $\mu = \sqrt{\frac{1}{4}+(2D/\rho^2)}$, $\nu_m = -\sqrt{\varepsilon_m}$, $G_{n,m}$ is defined in (12) as $G_n(\nu_m)$, $z$ and $\theta$ are defined in (20a) and (20b), respectively. The integer $N_m$ is equal to $\left\lfloor \frac{1}{2}\left\{-\sqrt{\frac{1}{4}+(2D/\rho^2)}+\sqrt{\varepsilon_m}-1\right\}\right\rfloor$.

Now, the physical properties of the system are obtained from those of the associated TRA polynomials $H_n^{(\mu,\nu)}(z;\gamma,\theta)$ and $\tilde{H}_n^{(\mu,\nu)}(z;\gamma,\theta)$. Here again and due to the lack of knowledge of the complete analytic properties of these TRA polynomials, we resort to numerical means to determine the energy spectrum as explained in Appendix B. Table 2 gives the lowest energy spectrum for two sets of potential parameters using two independent numerical routines. With the energy eigenvalues being found, the complete wavefunction is now fully determined as given by (21). Figure 4 is a plot of the un-normalized wavefunctions corresponding to the lowest energy eigenvalues in Table 2.

## 4. Conclusion and Discussion

This work is an additional testimony to the power and viability of the tridiagonal representation approach in solving novel quantum mechanical problems. The approach is endowed with robust mathematical underpinnings for solving ordinary second order differential equations of which the Schrödinger equation is but an example. The advantage of the approach in being algebraic is reinforced by the analytic power of orthogonal polynomials and special functions. On the computational side, it is favored as being reliant on powerful numerical techniques that deal with tridiagonal matrices and Gauss quadrature. Here, we presented bound state solutions for two generalizations of the trigonometric and hyperbolic Pöschl-Teller potentials, which are known to have a wide range of applications in atomic, molecular and nuclear physics. The generalized models are believed to have similar characteristics and fruitful applications.

It should be obvious from the structure of the hyperbolic potential $V_I(x)$, which is also visually evident in Fig. 1a, that it can be used to give an improved modeling of molecular binding that could be adjusted using four parameters $\{V_0, A, B, \kappa\}$ instead of the usual two or three potential parameters. Additionally, with a proper choice of these parameters, the potential may have local minimum and maximum that could place the system in the green phase of the SPD with a mix of bound states and resonances, which enriches the model depending on the value of the energy barrier measured by the difference between the two potential extrema $\Delta V$. On the other hand, the trigonometric potential $V_{II}(x)$, could be used in nuclear and particle physics as an alternative confining potential model (with a 4-parameter flexibility control) to the frequently used square potential box or the 1D Woods-Saxon potential well.

Nonetheless, a shortcoming of the TRA in the current problem is the absence of knowledge of the full analytic properties of the polynomials $H_n^{(\mu,\nu)}(z;\gamma,\theta)$ and $\tilde{H}_n^{(\mu,\nu)}(z;\gamma,\theta)$. These analytic properties contain all physical features of the system such as the energy spectrum. On the other hand, due to the typically small number of terms in the solution series (14) and (21), these polynomials could be evaluated explicitly using their three-term recursion relation and initial values. However, we had to obtain the energy spectrum by numerical means.

Due to the finiteness of the energy spectrum of the potential $V_I(x)$, the finite series (14) should give a faithful representation of the solution. However, since the energy spectrum of the potential $V_{II}(x)$ is infinite, the finite series (21) represents a quasi-exact solution for the lowest part of the spectrum whose accuracy could be improved by increasing the number of terms in the series (i.e., increasing the size of the basis).

## Appendix A: Jacobi polynomial on the semi-infinite line

These polynomials are defined over the semi-infinite interval $y \geq 1$ whereas the conventional Jacobi polynomials are defined on the finite interval $-1 \leq y \leq +1$. To distinguish it from the conventional Jacobi polynomials $P_n^{(\mu,\nu)}(y)$ we use the notation $Q_n^{(\mu,\nu)}(y)$:

$$Q_n^{(\mu,\nu)}(y) = \frac{\Gamma(n+\mu+1)}{\Gamma(n+1)\Gamma(\mu+1)}\, {}_2F_1\left(\begin{matrix}-n, n+\mu+\nu+1 \\ \mu+1\end{matrix}\bigg|\frac{1-y}{2}\right) = (-1)^n Q_n^{(\nu,\mu)}(-y). \tag{A1}$$

where $n = 0,1,2,...,N$, $\mu > -1$ and $\mu+\nu < -2N-1$. It satisfies the following differential equation

$$\left\{(y^2-1)\frac{d^2}{dy^2}+\left[(\mu+\nu+2)y+\mu-\nu\right]\frac{d}{dy}-n(n+\mu+\nu+1)\right\}Q_n^{(\mu,\nu)}(y) = 0, \tag{A2}$$

It also satisfies the following three-term recursion relation

$$y Q_n^{(\mu,\nu)}(y) = \frac{\nu^2-\mu^2}{(2n+\mu+\nu)(2n+\mu+\nu+2)}Q_n^{(\mu,\nu)}(y)$$
$$+\frac{2(n+\mu)(n+\nu)}{(2n+\mu+\nu)(2n+\mu+\nu+1)}Q_{n-1}^{(\mu,\nu)}(y)+\frac{2(n+1)(n+\mu+\nu+1)}{(2n+\mu+\nu+1)(2n+\mu+\nu+2)}Q_{n+1}^{(\mu,\nu)}(y) \tag{A3}$$

and the following differential relation

$$(y^2-1)\frac{d}{dy}Q_n^{(\mu,\nu)} = 2(n+\mu+\nu+1)\left[\frac{(\nu-\mu)n}{(2n+\mu+\nu)(2n+\mu+\nu+2)}Q_n^{(\mu,\nu)}\right.$$
$$\left.-\frac{(n+\mu)(n+\nu)}{(2n+\mu+\nu)(2n+\mu+\nu+1)}Q_{n-1}^{(\mu,\nu)}+\frac{n(n+1)}{(2n+\mu+\nu+1)(2n+\mu+\nu+2)}Q_{n+1}^{(\mu,\nu)}\right] \tag{A4}$$

The associated orthogonality relation reads as follows

$$\int_1^\infty (y-1)^\mu(y+1)^\nu Q_n^{(\mu,\nu)}(y)Q_m^{(\mu,\nu)}(y)dy = \frac{2^{\mu+\nu+1}}{2n+\mu+\nu+1}\frac{\Gamma(n+\mu+1)\Gamma(n+\nu+1)}{\Gamma(n+1)\Gamma(n+\mu+\nu+1)}\frac{\sin\pi\nu}{\sin\pi(\mu+\nu+1)}\delta_{nm}, \tag{A5}$$

where $n,m \in \{0,1,2,...,N\}$. Equivalently (see Eq. 4.9 in Ref. [19]),

$$\int_1^\infty (y-1)^\mu(y+1)^\nu Q_n^{(\mu,\nu)}(y)Q_m^{(\mu,\nu)}(y)dy = \frac{(-1)^{n+1}2^{\mu+\nu+1}}{2n+\mu+\nu+1}\frac{\Gamma(n+\mu+1)\Gamma(n+\nu+1)\Gamma(-n-\mu-\nu)}{\Gamma(n+1)\Gamma(-\nu)\Gamma(\nu+1)}\delta_{nm}. \tag{A6}$$

The polynomial $H_n^{(\mu,\nu)}(z;\alpha,\theta)$ is defined in Ref. [16] by its three-term recursion relation, which is Eq. (8) therein that reads

$$\left(\cos\theta\right)H_n^{(\mu,\nu)}(z;\gamma,\theta)=\left\{\left[\left(n+\tfrac{\mu+\nu+1}{2}\right)^2-\gamma^2\right]z\sin\theta+\tfrac{\nu^2-\mu^2}{(2n+\mu+\nu)(2n+\mu+\nu+2)}\right\}H_n^{(\mu,\nu)}(z;\gamma,\theta)$$
$$+\tfrac{2(n+\mu)(n+\nu)}{(2n+\mu+\nu)(2n+\mu+\nu+1)}H_{n-1}^{(\mu,\nu)}(z;\gamma,\theta)+\tfrac{2(n+1)(n+\mu+\nu+1)}{(2n+\mu+\nu+1)(2n+\mu+\nu+2)}H_{n+1}^{(\mu,\nu)}(z;\gamma,\theta),\tag{A7}$$

where $H_0^{(\mu,\nu)}(z;\alpha,\theta)=1$ and $H_{-1}^{(\mu,\nu)}(z;\alpha,\theta):=0$. For some range of values of the polynomial parameters, it is more appropriate to define $\tilde{H}_n^{(\mu,\nu)}(z;\gamma,\theta)=H_n^{(\mu,\nu)}(-iz;\alpha,i\theta)$, which maps the recursion (A7) into

$$\left(\cosh\theta\right)\tilde{H}_n^{(\mu,\nu)}(z;\gamma,\theta)=\left\{\left[\left(n+\tfrac{\mu+\nu+1}{2}\right)^2-\gamma^2\right]z\sinh\theta+\tfrac{\nu^2-\mu^2}{(2n+\mu+\nu)(2n+\mu+\nu+2)}\right\}\tilde{H}_n^{(\mu,\nu)}(z;\gamma,\theta)$$
$$+\tfrac{2(n+\mu)(n+\nu)}{(2n+\mu+\nu)(2n+\mu+\nu+1)}\tilde{H}_{n-1}^{(\mu,\nu)}(z;\gamma,\theta)+\tfrac{2(n+1)(n+\mu+\nu+1)}{(2n+\mu+\nu+1)(2n+\mu+\nu+2)}\tilde{H}_{n+1}^{(\mu,\nu)}(z;\gamma,\theta),\tag{A8}$$

## Appendix B: Energy spectrum and SPD

In this Appendix, we present two independent numerical techniques to calculate the energy spectrum and give a brief description of the spectral phase diagram (SPD) for the hyperbolic potential $V_{\mathrm{I}}(x)$.

### B.1 Discrete variable representation (DVR)

In the DVR, the 1D Hamiltonian, $H=-\tfrac{1}{2}\tfrac{d^2}{dx^2}+V(x)$, is represented in a given basis on a discrete grid. The choice of basis functions can take different forms [20-24]. In this work, we choose the particle-in-a-box basis, where the Hamiltonian is represented as (refer to [24] for details)

$$H_{ij}=\left(-\frac{1}{2}\frac{d^2}{dx^2}\right)_{ij}+V(x_i)\delta_{ij},\tag{B1}$$

where

$$\left(-\frac{1}{2}\frac{d^2}{dx^2}\right)_{ij}=\frac{(-1)^{i-j}}{(\Delta x)^2}\begin{cases}\dfrac{\pi^2}{6}-\dfrac{1}{4i^2} & ,i=j\\[6pt]\dfrac{1}{(i-j)^2}-\dfrac{1}{(i+j)^2} & ,i\neq j\end{cases}\tag{B2}$$

for problems defined on the semi-infinite interval $(0,\infty)$ with $\Delta x=b/M$ for some large enough cell size $b$ and $M$ is the size of the grid ($x_i=ib/M$ and $i=1,2,...,M-1$). For the hyperbolic potential $V_{\mathrm{I}}(x)$, convergence of the results for the chosen accuracy was reached with a grid size $M=200$ and for $b=10$. The full energy spectrum for two sets of potential parameters are shown in Table 1

For the trigonometric potential $V_{\mathrm{II}}(x)$, we used the following discrete representation of the kinetic energy term [24]

$$\left(-\frac{1}{2}\frac{d^2}{dx^2}\right)_{ij} = \frac{(-1)^{i-j}}{a^2}\frac{\pi^2}{4}\begin{cases}\frac{2M^2+1}{3}-\frac{1}{\sin^2(i\pi/M)} & ,i=j \\ \frac{1}{\sin^2\left(\frac{i-j}{2M}\pi\right)}-\frac{1}{\sin^2\left(\frac{i+j}{2M}\pi\right)} & ,i\neq j\end{cases} \quad (B3)$$

We have studied the convergence of the lowest 10 eigenvalues of the trigonometric potential versus the grid size and reached stable results for the chosen accuracy with $M=300$ and $a=1$. The lowest part of the energy spectrum with two sets of physical parameters are shown in Table 2. We should note that both expressions for the kinetic energy operator are related where (B2) can be obtained from (B3) by taking the limits $a,M\to\infty$ such that $\Delta x=a/M$ is finite [24].

**B.2 Higher order finite difference (HOFD)**

Mebirouk *et al* [25] introduced a computational technique for problems defined on the semi-infinite interval $x\in(0,\infty)$ by mapping the configuration space as $x\mapsto s(x)$ where $s\in(0,1)$ which can then be easily studied via the higher order finite difference (HOFD) schemes. The transformation used was $s=\frac{2}{\pi}\tan^{-1}(\zeta x)$, where $\zeta=0.6(j)^{-0.7}$ for $j=1,2,...,M$ is the eigenvalue index for grid size $M$ (for details refer to [25]). This scheme was applied to the hyperbolic potential $V_1(x)$, where the associated Schrödinger equation is mapped into the following second order differential equation

$$\frac{2\zeta^2}{\pi^2}\left\{-\cos^4\left(\tfrac{\pi}{2}s\right)\frac{d^2}{ds^2}+\pi\cos^3\left(\tfrac{\pi}{2}s\right)\sin\left(\tfrac{\pi}{2}s\right)\frac{d}{ds}+\frac{\pi^2}{2\zeta^2}[V_1(s)-E]\right\}\psi(s)=0. \quad (B4)$$

The derivatives are replaced by the following approximations

$$\left.\frac{d\psi}{ds}\right|_{s_i}\approx\frac{1}{h}\begin{cases}\sum_{j=0}^{2k}\delta_{i,j,1}\psi_j, & i=1,2,3,...,k-1 \\ \sum_{j=0}^{2k}\delta_{k,j,1}\psi_{i+j-k}, & i=k,k+1,...,M+1-k \\ \sum_{j=0}^{2k}\delta_{i-t,j,1}\psi_{j+t}, & i=M+2-k,...,M\end{cases} \quad (B5a)$$

$$\left.\frac{d^2\psi}{ds^2}\right|_{s_i}\approx\frac{1}{h^2}\begin{cases}\sum_{j=0}^{2k+1}\delta_{i,j,2}\psi_j, & i=1,2,3,...,k-1 \\ \sum_{j=0}^{2k}\delta_{k,j,2}\psi_{i+j-k}, & i=k,k+1,...,M+1-k \\ \sum_{j=0}^{2k+1}\delta_{k,j,2}\psi_{j+t-1}, & i=M+2-k,...,M\end{cases} \quad (B5b)$$

where $t=M-2k+1$ and $\{\delta_{i,j,l}\}$ with $l=1,2$ are the weights obtained by requiring maximum order of consistency [25]. We used a uniform grid as $s_i=i/(M+1)$, where $i=0,1,...,M+1$. Using the boundary condition $\psi_0=\psi_{M+1}=0$ for Eqs. (B5) in Eq. (B4), we get

$$J\Psi = \left(\tilde{A}\Delta_2 + \tilde{B}\Delta_1 + \tilde{C}I_M\right)\Psi = E\Psi, \tag{B6}$$

where $\tilde{A}$, $\tilde{B}$ and $\tilde{C}$ are diagonal matrices of size $M$ which contain the coefficients of the Schrödinger wave operator (B4). $\Delta_l$ with $l = 1, 2$ are real square matrices of size $M$ containing the weights $\delta_{i,j,l}/h^l$, $I_M$ is the $M \times M$ identity matrix and $\Psi = (\psi_1, \psi_2, \ldots, \psi_M)^T$. Then, the eigenvalues and the corresponding eigenvectors of the matrix $J$ determines the complete solutions of the Schrödinger equation. We compare the eigenvalues obtained via the HOFD versus those obtained using the DVR as shown in Table 1 where we observe a very good agreement between the two method.

On the other hand, for the trigonometric potential $V_{II}(x)$, the problem is much simpler as we are dealing with the finite interval $(0, a)$ where no coordinate transformation is needed, just a simple rescaling to the interval $(0,1)$. One can also replace the second derivative in the wave equation by Eq. (B5b) above. Our results with the HOFD are given in Table 2 where we compare them against those obtained via the DVR. Again, a good agreement between the two methods is evident.

**B.3 Spectral phase diagram (SPD)**

Assi *et al* [11] introduced the concept of a spectral phase diagram (SPD) for non-confining potentials. The diagram shows the different phases that the system can exist in for various choice of parameters. The SPD is generally divided into four regions: Bound states, mix of bound and resonance states, resonances only, and scattering states only. Those regions are based on the necessary (may not be sufficient) conditions for the potential to support any of these states.

The SPD is useful for choosing values of the potential parameters to place the system in a certain spectral phase. Clearly, the SPD for the potential well $V_{II}(x)$ is trivial where it has only one region as it only supports bound states. Thus, we have generated the SPD only for $V_I(x)$ as given in Figure 2. In the figure, "B" stands for bound states (blue color), "R" stands for resonance states (red color), "B&R" stands for regions where a mix of bound states and resonances exist (green color), and "S" where *only* scattering states can exist (grey color). Note, of course, that scattering states occur in all regions of the SPD. In the figure we show the TRA solution space as the rectangular area indicated by the TRA constraints $B \leq \kappa^2/8$ and $A < V_0$. It is also clear from Fig. 2 that a necessary condition for pure bound states is that $(A + B) < 0$, which is consistent with the observation made below Eq. (2) above.

# Acknowledgements

We are grateful to Prof. H. Bahlouli (KFUPM) for fruitful discussions and for the careful review and improvements of the initial version of the manuscript.

## Tables Captions

**Table 1.** The complete energy spectrum (in atomic units) for the hyperbolic potential $V_I(x)$ obtained using DVR and HOFD techniques with two sets of potential parameters: $S_1 = \{V_0, A, B, \kappa\} = \{10, -20, -30, 1\}$ and $S_2 = \{V_0, A, B, \kappa\} = \{5, 2, -60, 1\}$. For the DVR, we took $b = 10$ and a grid size $M = 200$ whereas for the HOFD, we took $M = 500$.

**Table 2.** The lowest part of the energy spectrum (in atomic units) for the trigonometric potential $V_{II}(x)$ obtained using DVR and HOFD techniques with two sets of potential parameters: $S_3 = \{V_0, C, D, a\} = \{5, -10, 2, 1\}$ and $S_4 = \{V_0, C, D, a\} = \{5, -2, 2, 1\}$. For the DVR, we took a grid size $M = 300$ whereas for the HOFD, we took $M = 500$.

## Figures Captions

**Fig. 1.** The two potential functions $V_I(x)$ and $V_{II}(x)$ (in units of $V_0$) for several values of the associated parameters: (a) $V_I(x)$ with $B = 2V_0$ and several values of $A$, (b) $V_{II}(x)$ with $D = 2V_0$ and several values of $C$.

**Fig. 2.** The SPD for the hyperbolic potential $V_I(x)$ (see section B.3 in Appendix B for description). The TRA bound-state solution space is indicated by the interior of the rectangular area bounded by $B = \kappa^2/8$ and $A = V_0$.

**Fig. 3.** The un-normalized wavefunctions corresponding to the full energy spectrum of $V_I(x)$ in Table 1. The top row corresponds to the parameter set $S_1$ and the bottom row corresponds to the parameter set $S_2$. The horizontal axis is $\kappa x$.

**Fig. 4.** The un-normalized wavefunctions corresponding to the lowest energy spectrum of $V_{II}(x)$ in Table 2. The top row corresponds to the parameter set $S_3$ and the bottom row corresponds to the parameter set $S_4$. The horizontal axis is $x/a$.

**Table 1**

|   | $S_1$ | | $S_2$ | |
|---|---|---|---|---|
| $n$ | DVR | HOFD | DVR | HOFD |
| 0 | −17.292792568552 | −17.292792568575 | −15.992869980420 | −15.992869980437 |
| 1 | −6.137201742096 | −6.137201742113 | −6.101528843700 | −6.101528843717 |
| 2 | −0.888027613576 | −0.888027616853 | −1.000393053814 | −1.000393054957 |

**Table 2**

|   | $S_3$ | | $S_4$ | |
|---|---|---|---|---|
| $n$ | DVR | HOFD | DVR | HOFD |
| 0 | 16.797026 | 16.797032 | 29.961374 | 29.961382 |
| 1 | 53.186883 | 53.186917 | 68.685118 | 68.685159 |
| 2 | 103.396936 | 103.397040 | 120.819954 | 120.820074 |
| 3 | 166.730521 | 166.730761 | 185.823763 | 185.824031 |
| 4 | 242.759201 | 242.759670 | 263.346993 | 263.347504 |
| 5 | 331.187625 | 331.188444 | 353.139727 | 353.140605 |
| 6 | 431.796715 | 431.798037 | 455.011712 | 455.013113 |
| 7 | 544.415737 | 544.417750 | 568.811809 | 568.813926 |
| 8 | 668.906827 | 668.909756 | 694.416181 | 694.419241 |
| 9 | 805.155660 | 805.159769 | 831.720941 | 831.725211 |

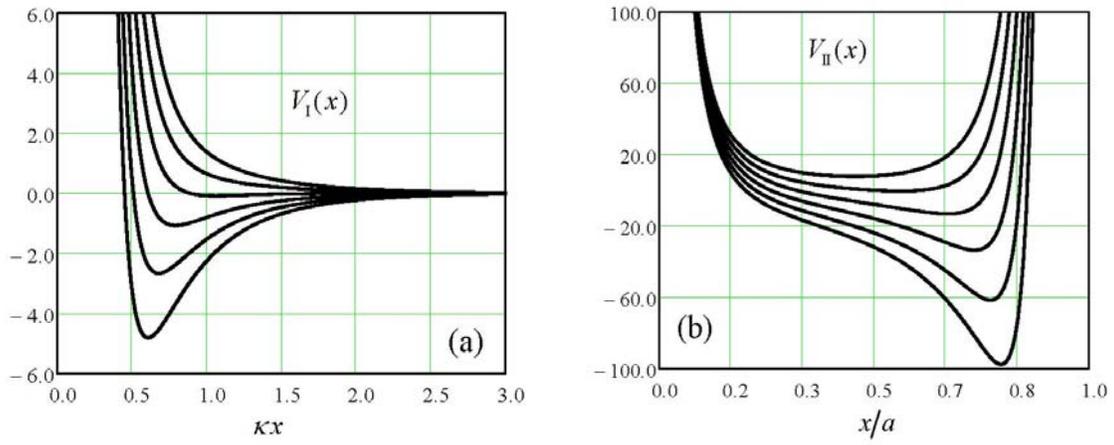

**Fig. 1**

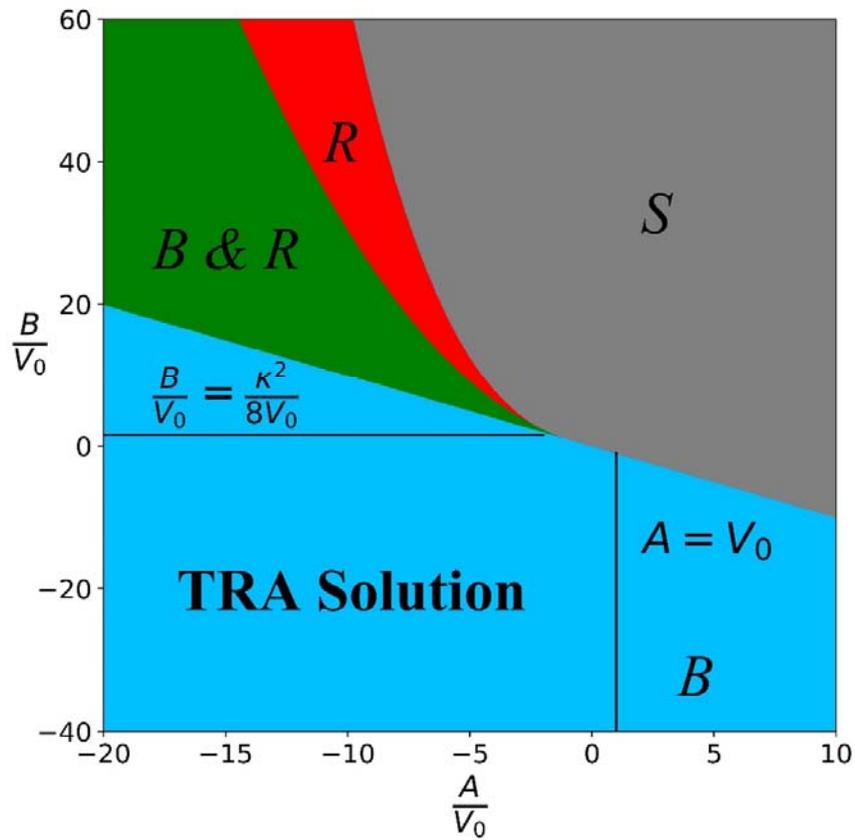

**Fig. 2**

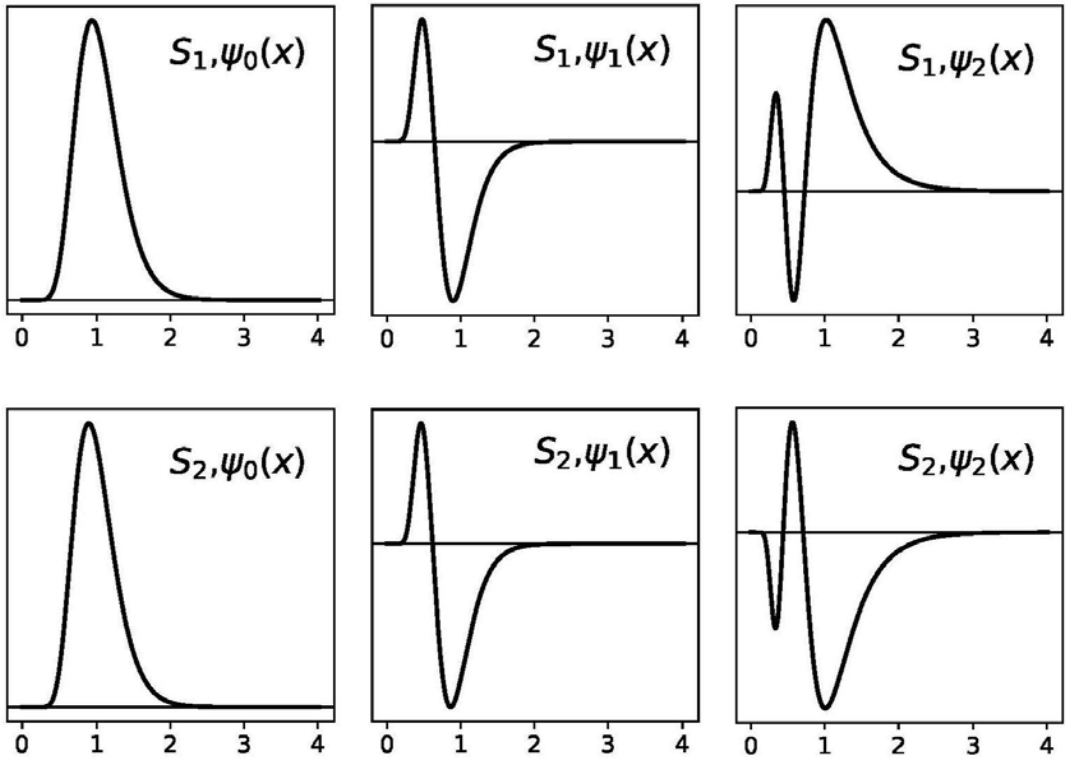

**Fig. 3**

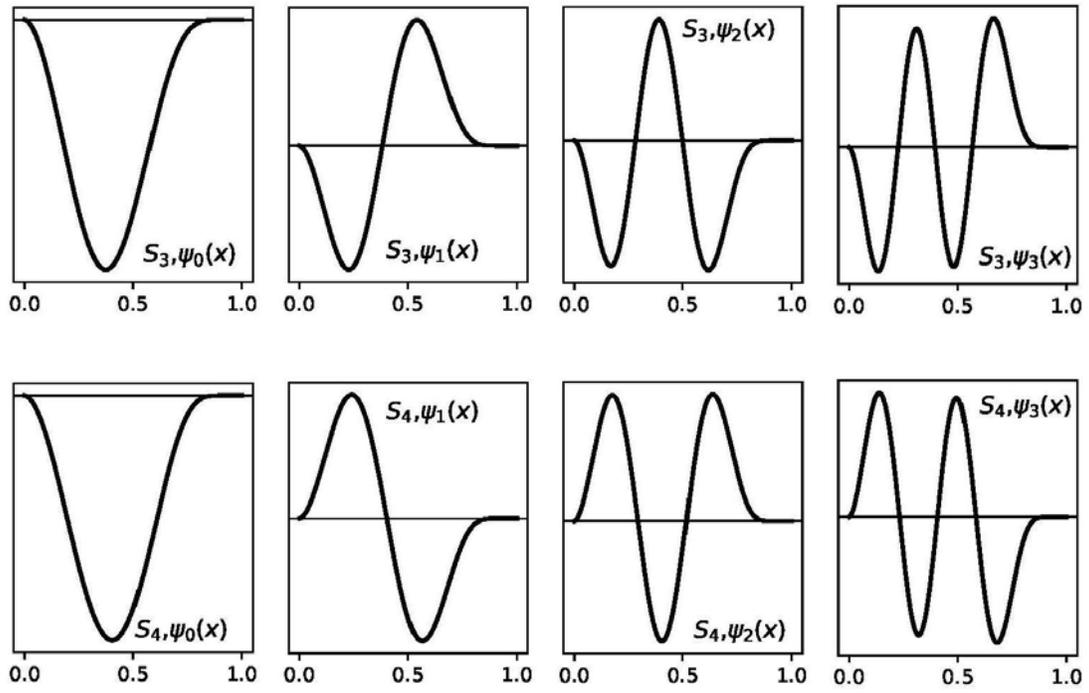

**Fig. 4**